\documentclass[11pt]{article}
\usepackage{moriond,epsfig}
\input psfig.tex 

\def\PRD{{\em Phys. Rev.} D}

\def\be{\begin{equation}}
\def\ee{\end{equation}}
\def\bea{\begin{eqnarray}}
\def\eea{\end{eqnarray}}

\begin{document}
\vspace*{4cm}
\title{PROBING THE PRIMORDIAL POWER SPECTRUM WITH THE VSA \\(AND OTHER CMB EXPERIMENTS)}

\author{Richard A. Battye and the VSA Collaboration}

\address{Jodrell Bank Observatory, Department of Physics and Astronomy,\\
University of Manchester, Macclesfield SK11 9DL, U.K.}

\maketitle\abstracts{We present estimates of parameters which model the primordial density fluctuations using CMB data from the VSA, 1st year WMAP, CBI and ACBAR along with external priors from 2dF and the HST. Taken at face value there is significant evidence for a running spectral index when only scalar fluctuations are considered. The significance of this result is increased by the inclusion of tensors and, although the result is weakened by the inclusion of `running of the running', the basic trend toward significantly negative running remains. We discuss the implications of this result and how it might be strengthened by future observations.}

\section{Introduction}

The Very Small Array (VSA) is a 14 element interferometer which has been operating in the Ka band ($\nu\approx 26-36{\rm GHz}$) at the Teide Observatory in Tenerife since 2000. The present discussion concerns the most recently presented  measurements~\cite{p7,p8} which are discussed elsewhere in these proceedings~\cite{keith}. They represent $\sim 7000$ hours of actual integration on a total of 33 fields by the VSA in its extended  configuration. The specific results presented here focus primarily on the constraints obtained on the primordial power spectrum in conjunction with the 1st year WMAP data~\cite{WMAP}, the most recent CBI results~\cite{CBI} and those from ACBAR~\cite{ACBAR}. They follow up and expand upon those already reported~\cite{p8}.

We use the COSMOMC package~\cite{COSMOMC} to perform our calculations and  maginalize over the other cosmological parameters $\Omega_{\rm b}h^2$, $\Omega_{\rm m}h^2$, $h$, $A_{\rm S}$ and $z_{\rm re}$ as described in previous work~\cite{p8}. We use the same flat priors except for the redshift of reionization, $z_{\rm re}$, for which we use the range $(4,50)$.

The scalar component of the primordial power spectrum is modelled as a power series in $\log(k/k_{\rm c})$, 
\begin{equation}
\log P_{\rm S}(k)=\log A_{\rm S} + (n_{\rm S}-1)\log (k/k_{\rm c}) +{1\over 2}n_{\rm R}[\log(k/k_{\rm c})]^2+{1\over 6}n_{\rm RR}[\log(k/k_{\rm c})]^3+.. 
\end{equation}
where $k_{\rm c}$ is the pivot scale, $A_{\rm S}$ is the amplitude of the power spectrum, $n_{\rm S}$ is the spectral index, $n_{\rm R}$ ($n_{\rm run}$ in ref.[2]) is the running of the spectral index and $n_{\rm RR}$ is the running of the running of the spectral index. For this to be considered a valid power series expansion, one will require each of the terms to be smaller than the previous one over the range of $k$ relevant to cosmological observations $(\log(k/k_{\rm c})\approx 4)$. This requires that $|n_{\rm S}-1|\gg 2n_{\rm R}\gg 8n_{\rm RR}/3$.

We will consider three different models (A) Scalar fluctuations only with $n_{\rm RR}=0$; (B) Scalar fluctuations with $n_{\rm RR}=0$ and also tensor fluctuations with power spectrum $P_{\rm T}(k)=A_{\rm T}(k/k_{\rm c})^{n_{\rm T}}$; (C) Scalar fluctuations only, but with $n_{\rm RR}\ne 0$. For models A and C we will use $k_{\rm c}=0.05\,{\rm Mpc}^{-1}$ and for B, $k_{\rm c}=0.002\,{\rm Mpc}^{-1}$. Since we find that models with significant running are favoured, the preferred values of $n_{\rm S}$ will appear to be very different for model B when compared to A and C.

In addition to data from CMB experiments, we will use information from external sources to augment our analysis. In particular, we will use information from the HST Key Project~\cite{freed} on $h=0.72\pm 0.08$ with the errorbars assumed to be Gaussian (denoted HST) and from the 2dF Galaxy Redshift survey~\cite{percival} which constrains the matter power spectrum (denoted 2dF)

\begin{table}[t]
\caption{Estimates of $n_{\rm S}$, $n_{\rm R}$, $n_{\rm RR}$, $n_{\rm T}$ and  $A_{\rm T}/A_{\rm S}$ for models A, B and C. We have used CMB data from WV and WACV with external priors from 2dF and HST.}
\vspace{0.4cm}
\begin{center}
\begin{tabular}{|c|c|c|c|c|c|c|c|}
\hline\hline
& CMB  & EXT & $n_{\rm S}$ & $n_{\rm R}$ & $n_{\rm RR}$ & $n_{\rm T}$ & $A_{\rm T}/A_{\rm S}$ \\
\hline\hline
A & WV & HST & $0.93\pm 0.06$ & $-0.08 \pm 0.04$ & 0 & 0 & 0 \\
 & & 2dF & $0.93\pm 0.05$ & $-0.05\pm 0.03$ & 0 & 0 & 0  \\
& WACV & HST & $0.91\pm 0.06$ & $-0.08\pm 0.04$ & 0 & 0 & 0 \\
&  & 2dF & $0.90\pm 0.05$ & $-0.06\pm 0.03$ & 0 & 0 & 0 \\
\hline
B & WACV & HST & $1.26\pm 0.12$  & $-0.11\pm 0.04$ & 0 & $-0.02\pm 0.49$ & $0.49\pm 0.39$ \\
& & 2dF & $1.18\pm 0.11$  & $-0.08\pm 0.04$ & 0 & $-0.08\pm 0.42$ & $0.27\pm 0.22$ 
\\
\hline 
C & WACV & HST & $0.89\pm 0.07$ & $-0.18\pm 0.10$ & $-0.04\pm 0.05$ & 0 & 0  \\
&  & 2dF & $0.89\pm 0.08$ & $-0.09\pm 0.09$ & $-0.01\pm 0.05$ & 0 & 0  \\
\hline\hline
\end{tabular}
\label{tab:nnn}
\end{center}
\end{table}

\begin{figure}[!h] 
\setlength{\unitlength}{1cm}
\centerline{\psfig{file=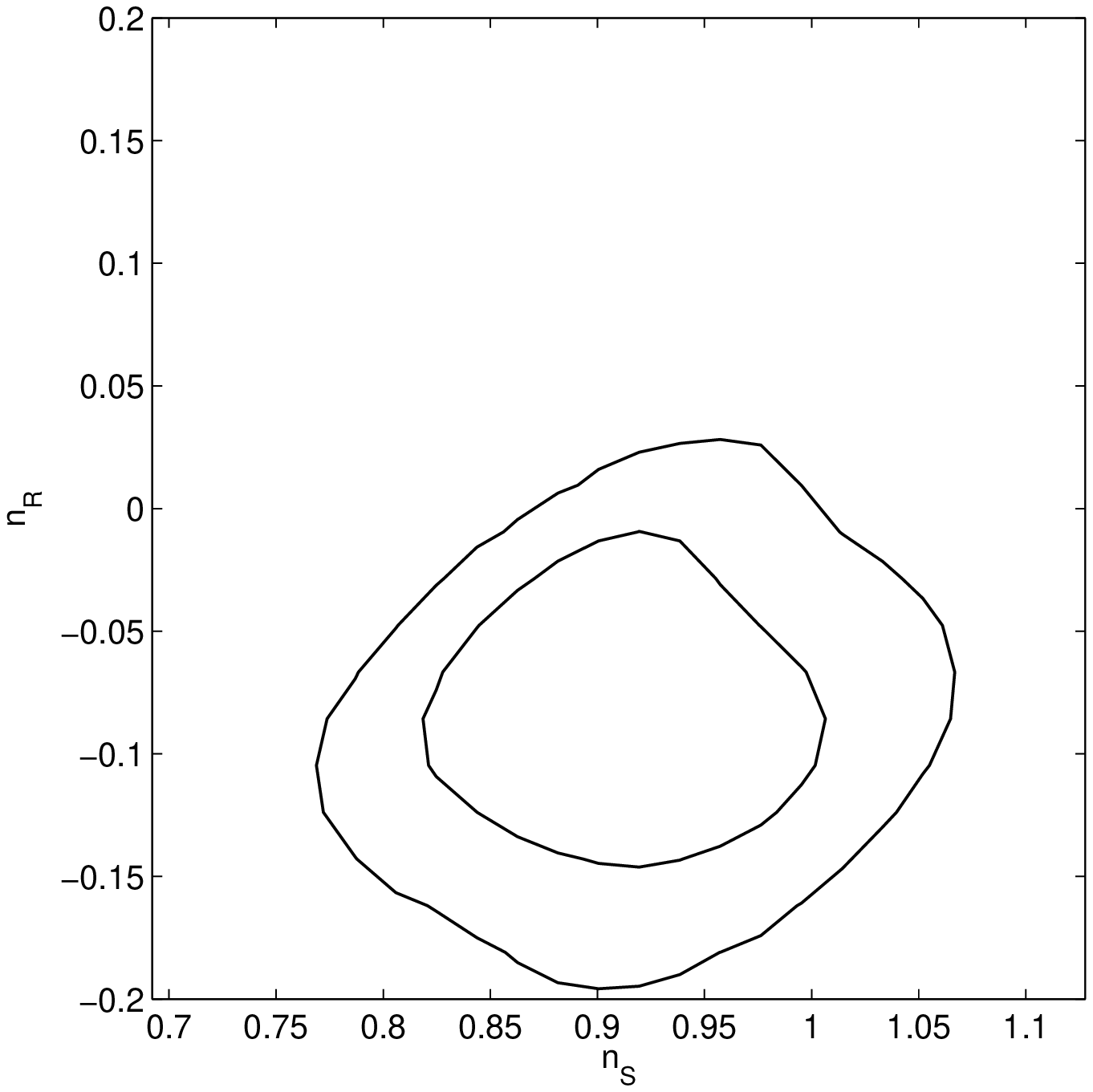,width=5cm,height=5cm}\qquad\qquad\psfig{file=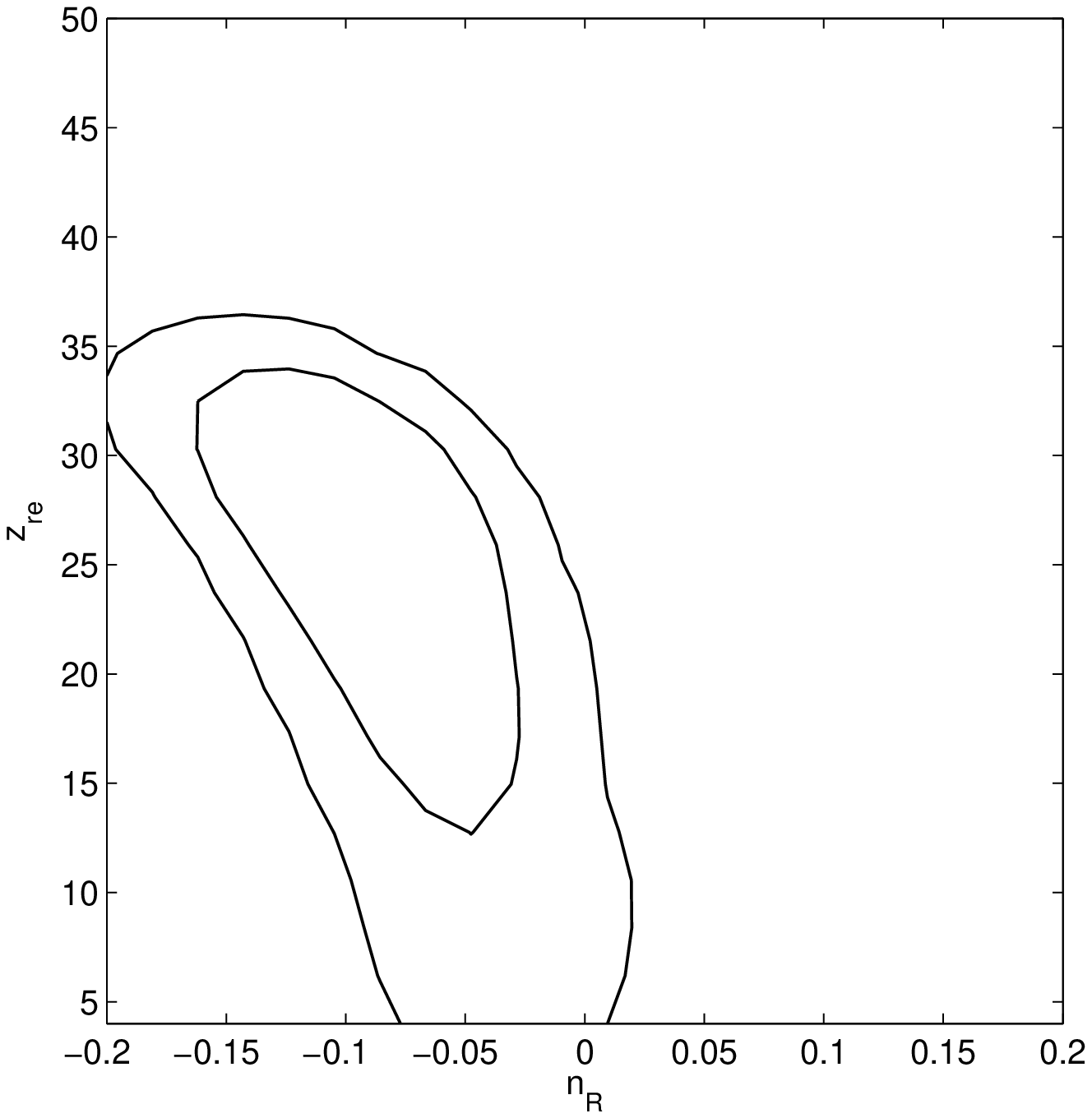,width=5cm,height=5cm}}
\caption{Marginalized 2D distributions for (a,left) $n_{\rm S}-n_{\rm R}$ and (b,right) $n_{\rm R}-z_{\rm re}$ for model A using WACV and HST. The contours represent the 68\% (inner) and 95\% (outer) confidence levels.} 
\label{fig:deg1}
\end{figure}	

\section{Results \& discussion}

The results of our analysis are presented in Table~\ref{tab:nnn} for models A,B and C for the CMB data from WV (WMAP+VSA) and WACV (WMAP+ACBAR+CBI+VSA), and the external priors. The most striking feature in the case of model A is that $n_{\rm R}<0$ is preferred and for the most part at $\ge2\sigma$. We see that when one uses the HST prior, the preferred value of $n_{\rm R}$ is lower than for the 2dF prior. We note that the preferred values of $n_{\rm R}$ are at odds with the conditions for the power series expansion to the valid, but not at the  $1\sigma$ level. Models with $n_{\rm R}<0$  have an absence of power on both large and small scales relative to scale invariance, with $n_{\rm S}>1$ on very large scales and $n_{\rm S}<1$ on very small scales. It appears that the data favour such models.

It is interesting to consider what slow-roll inflation predicts for $n_{\rm R}$. For a power-law model with $V(\phi)\propto\phi^p$, a simple approximation yields $n_{\rm S}(N)\approx 1-((p+2)/2N)$ ($\le1$ for all $N$) where $N$ is the number of e-foldings from the end of inflation when the corresponding comoving wavelength leaves the horizon. Hence, one can deduce that $n_{\rm R}\approx -2(n_{\rm S}-1)^2/(p+2)$ which the results clearly violate. In general all slow roll inflation models predict that $n_{\rm R}\sim -(n_{\rm S}-1)^2$ and our results present a significant challenge to them. Multi-field inflation models are much less well constrained by these results.

One bizarre feature of our results is that the preferred values of $z_{\rm re}$ are greater than one might expect, even in the most baroque models for reionization. For model A and WACV the constraint on $z_{\rm re}=23\pm7$ for the HST prior and $z_{\rm re}=18\pm 6$ for the 2dF prior. We have presented the degeneracies in the $n_{\rm R}-z_{\rm re}$ and $n_{\rm S}-n_{\rm R}$ planes for the HST prior in Fig.~\ref{fig:deg1}. We see that large negative values of $n_{\rm R}$ correspond to large value of $z_{\rm re}$. A unique feature of the present analysis is that we include the full range of values of $z_{\rm re}$ whereas various other analyses~\cite{p8,spergel} use restricted ranges. While excluding high values of $z_{\rm re}$ might be suggested by arguments about the nature of reionization, leaving them out of the analysis with a flat prior can bias the confidence limits on the other parameters by introducing a model dependent slicing of the likelihood surface. This issue will be resolved by upcoming WMAP polarization data which should constrain $z_{\rm re}$ much more tightly.

\begin{figure}[!h] 
\setlength{\unitlength}{1cm}
\centerline{\psfig{file=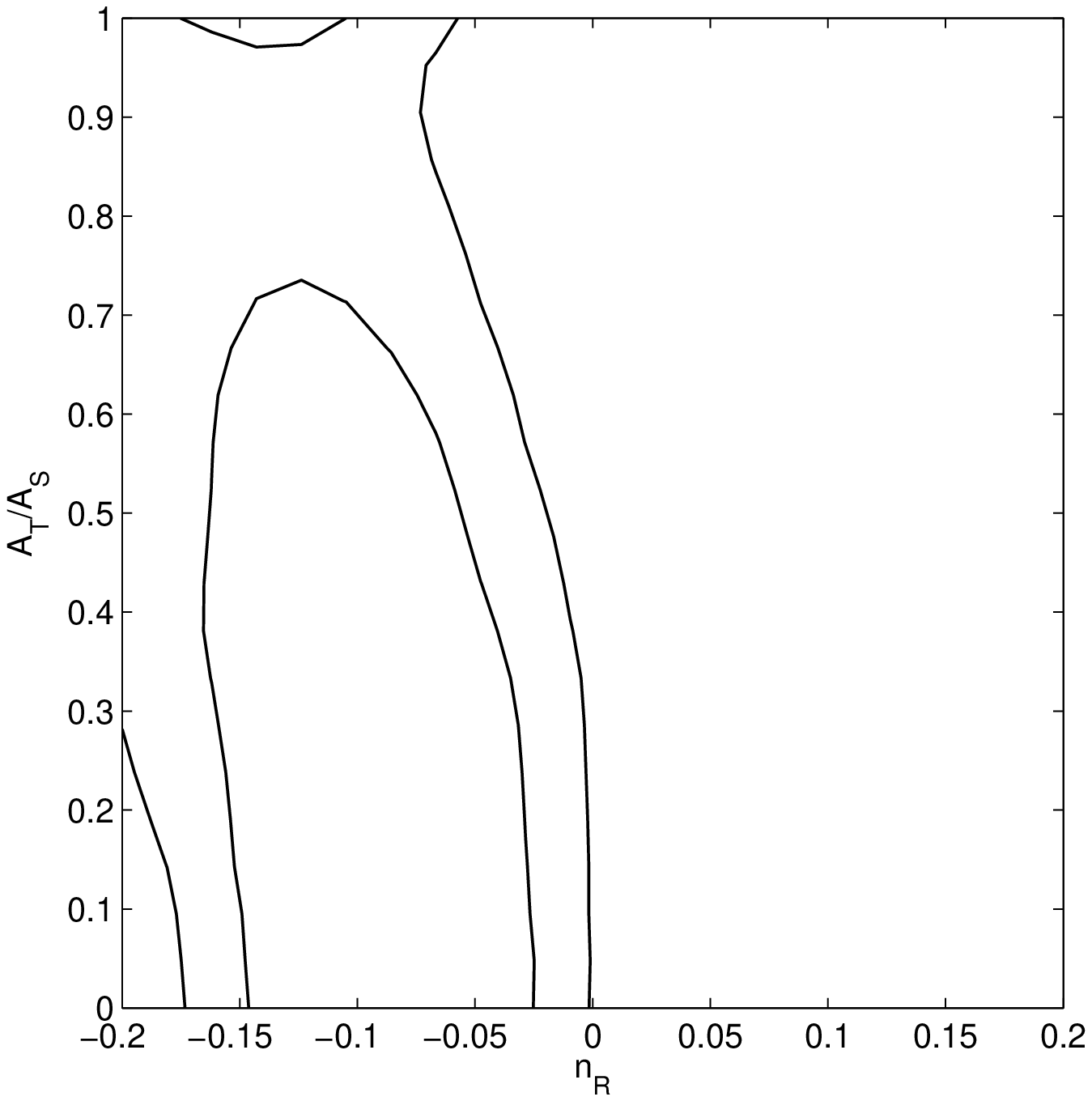,width=5cm,height=5cm}\qquad\qquad\psfig{file=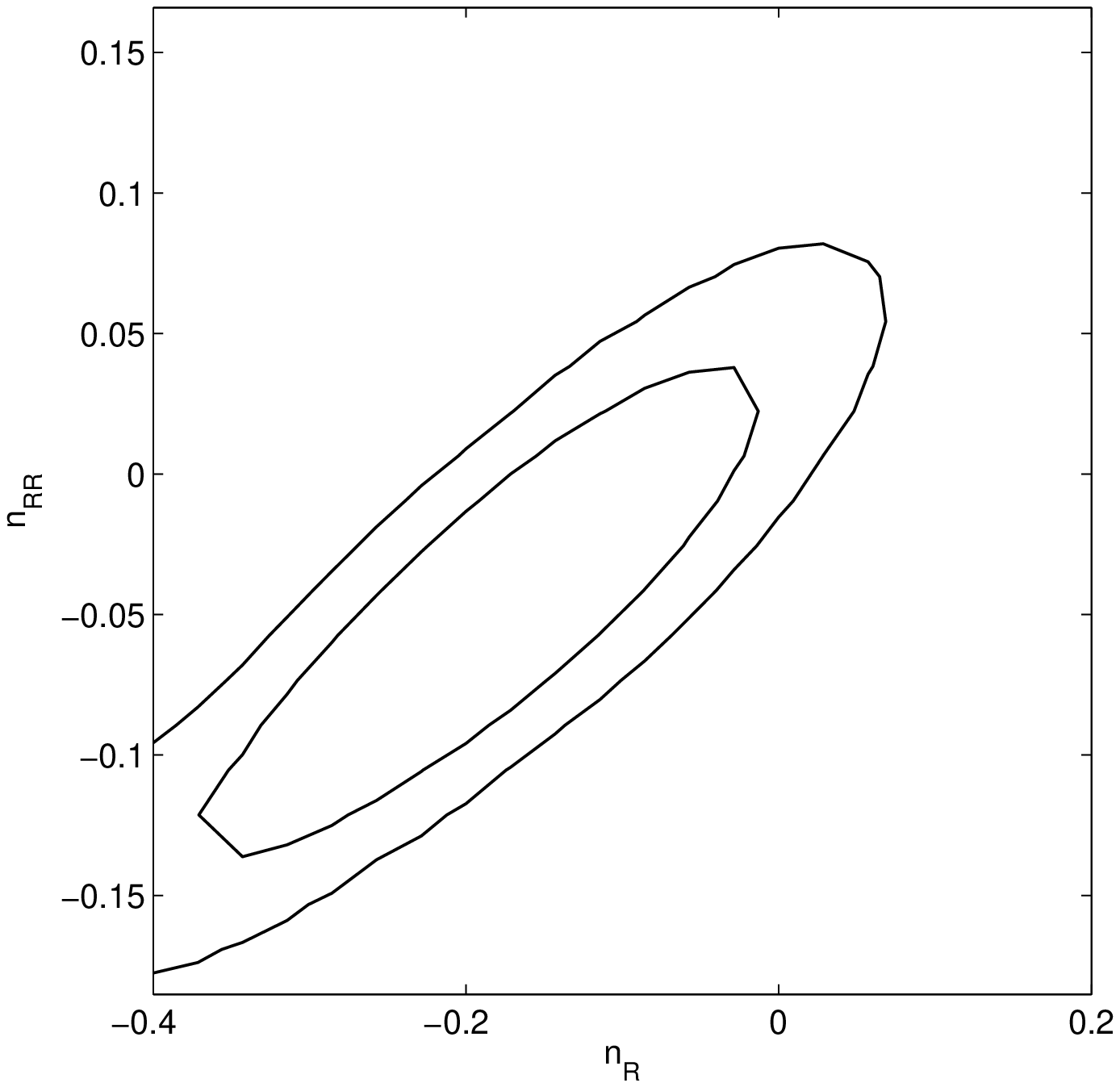,width=5cm,height=5cm}}
\caption{Marginalized 2D distributions for (a,left) $n_{\rm R}-A_{\rm T}/A_{\rm S}$ for model B using WACV and HST, and (b,right) $n_{\rm R}-n_{\rm RR}$ for model C using the same data. Confidence levels as in fig.~\ref{fig:deg1}.} 
\label{fig:deg2}
\end{figure}

A wide class of inflationary models predict a significant tensor amplitude. Our results for model B show that the preference for $n_{\rm R}<0$ is more significant when they are included in the analysis. This is because models with $A_{\rm T}/A_{\rm S}\ne 0$ prefer more negative values of $n_{\rm R}$ than for $A_{\rm T}/A_{\rm S}$ as illustrated in Fig.~\ref{fig:deg2}(a) for WACV and the HST prior.  

One might expect that if $n_{\rm R}$ is significantly $<0$ as suggested by our results then one might also expect that $n_{\rm RR}\ne 0$ also. Our results for model C appear to confirm the preference for $n_{\rm R}<0$, albeit at less significance,  while appearing to be consistent with $n_{\rm RR}=0$. There is a strong degeneracy between $n_{\rm R}$ and $n_{\rm RR}$ as illustrated in Fig.~\ref{fig:deg2}(b).

We should note our previous work~\cite{p8} highlighted a number of possible systematic effects which could be at work here. The parameter $n_{\rm R}$ can only be constrained by making measurements over a range of angular scales making it particularly sensitive to a wide range of effects. In particular, we highlighted issues pertaining to the overall calibration uncertainty of the data, possible errors in the point source subtraction, and the possibility that the measured WMAP quadrapole is contaminated by foreground emission.

Finally, in order to re-inforce the point that the data favour models with a dearth of power on small scales, we have also considered a model with a broken power law spectral index. Specifically, we have used $n_{\rm S}(k)=n_1$ for $k<k_{\rm c}$ and $n_{\rm S}=n_2$ for $k>k_{\rm c}$, with $k_{\rm c}=0.05\,{\rm Mpc}^{-1}$. We find that for WV+HST $n_1=1.05\pm 0.05$ and $n_2=0.54\pm 0.22$, whereas for  WV+2dF $n_1=0.98\pm 0.05$ and $n_2=0.87\pm 0.15$. Although this should not be taken too seriously as physical model, it illustrates the point very strongly.

\section{Future observations using the super-extended array}

It is planned that the VSA will be reconfigured in the late summer of 2004 and fitted with larger mirrors to achieve much higher angular resolution. It is expected that it will be possible to make observations which are sensitive to $\ell=500-2500$. Given that the primary CMB signal (in temperature) at $\ell=2000$ is a factor of about 3 weaker than at $\ell=1000$, it is planned to upgrade the array electronics to give a factor $\sim 3$ improvement in sensitivity to maintain the same signal-to-noise. This will be achieved by an increase in the bandwidth from $\Delta f=1.5\,{\rm GHz}$ to $6\,{\rm GHz}$ and a reduction in $T_{\rm sys}$ from 35K to 25K. 

We have produced simulated errorbars for an 18 month observing program and used them to create a mock COSMOMC file based on a fiducial model which is the best fit to WV+HST. We found that the derived error on the running of the spectral index  $\Delta n_{\rm R}$ reduces by about 30\% from $\Delta n_{\rm R}=0.038$ to $\Delta n_{\rm R}=0.027$ suggesting that observations at high $\ell$ can have a significant impact on the constraining the primordial power spectrum.

\section*{Acknowledgments}
I would like to thank my colleagues who work tirelessly on the operation of and analysis of data from the VSA. I am particularly indebted to Richard Davis, Clive Dickinson and Kieran Cleary. I am currently funded by a PPARC Advanced Fellowship and the funding for building and operation of the VSA was also provided from PPARC. I thank Anthony Lewis for making COSMOMC publically available. I apologise for the lack of comprehensive references which is necessitated by space constraints.


\section*{References}
\begin{thebibliography}{99}

\bibitem{p7}
C. Dickinson et al., astro-ph/0402498

\bibitem{p8}
R. Rebolo et al., astro-ph/0402466

\bibitem{keith}
K. Grainge, these proceedings.

\bibitem{WMAP}
C. Bennett et al, 2003, {\it ApJS} 148, 1

\bibitem{CBI}
A.C.S. Readhead et al, 2004, astro-ph/0402359

\bibitem{ACBAR}
C.L. Kuo et al, 2004, {\it ApJ} 600, 32

\bibitem{COSMOMC}
A. Lewis and S.L. Bridle, 2002, \PRD, 103511

\bibitem{freed}
W.L. Freedman et al, 2001, {\it ApJ} 553, 57

\bibitem{percival}
W.J. Percival et al, 2002, {\it MNRAS} 337, 1068

\bibitem{spergel}
D. Spergel er al, 2003, {\it ApJS} 148, 175

\end{thebibliography}
\end{document}